\documentclass[prb,aps,twocolumn,letter,showpacs,preprintnumbers,amsmath,amssymb]{revtex4}
\usepackage{graphicx}
\usepackage{dcolumn}
\usepackage{bm}
\usepackage{color}
\begin{document}
\newcommand*{\BiI}[0]{Bi$_2$Sr$_2$CuO$_{6 + \delta}$} 
\newcommand*{\BiII}[0]{Bi$_2$Sr$_2$CaCu$_2$O$_{8 + \delta}$} 
\newcommand*{\Tc}[0]{$T_{\rm c}$} 
\newcommand*{\Tcm}[0]{$T_{\rm c,max}$} 

\title{Effect of chemical inhomogeneity in the bismuth-based copper oxide superconductors}

\author{H.Eisaki}
 \altaffiliation[Present address: ]{Nanoelectronic Research Institute, AIST, Tsukuba 305-8568, 
Japan}
\author{N. Kaneko}
\altaffiliation[Present address: ]{National Metrology Institute of Japan, AIST, Tsukuba 305-8568, 
Japan}
\author{D.L. Feng }%
\altaffiliation[Present address: ]{Department of Physics \& Astronomy,
The University of British Columbia, 334-6224 Agricultural Rd. Vancouver, B.C. V6T 1Z1, Canada 
and Department of Physics, Fudan University, Shanghai, China}
\author{A. Damascelli }%
\altaffiliation[Present address: ]{Department of Physics \& Astronomy,
The University of British Columbia, 334-6224 Agricultural Rd. Vancouver, B.C. V6T 1Z1, Canada}
\author{P.K. Mang}
\author{K.M. Shen}
\author{Z.-X. Shen}
\author{M. Greven}

\affiliation{%
Department of Applied Physics, Physics, and Stanford Synchrotron Radiation Laboratory,
Stanford University, Stanford CA, 94305
}%

\date{\today}

\begin{abstract}

We examine the effect on the superconducting transition temperature (\Tc) of 
chemical inhomogeneities in \BiI~and 
\BiII~single crystals. Cation disorder at the Sr crystallographic 
site is inherent in these materials and strongly affects the value of \Tc. 
Partial substitution of Sr by Ln (Ln = La, Pr, Nd, Sm, Eu, Gd, and Bi) in
Bi$_2$Sr$_{1.6}$Ln$_{0.4}$CuO$_{6+ \delta}$ results in a monotonic decrease of \Tc~with
increasing ionic radius mismatch.
 By minimizing Sr site disorder at the expense of Ca site disorder, we demonstrate that 
the \Tc~of \BiII~can be increased to 96 K. Based on these results we discuss 
the effects of chemical inhomogeneity in other bulk high-temperature superconductors.
\end{abstract}


\maketitle

\section{INTRODUCTION}

The possible existence of nanoscale electronic inhomogeneity --- the propensity of 
charge carriers doped into the CuO$_2$ plane to form nanoscale structures --- has 
drawn much attention in the field of high-\Tc~superconductivity. 
Neutron scattering studies on Nd co-doped La$_{2-x}$Sr$_x$CuO$_4$ (Nd-LSCO)~\cite{neutron} 
and STM/STS studies on \BiII~(Bi2212)~\cite{STM} have led to suggestions that such 
self-organization may manifest itself as one-dimensional ``stripes'' in Nd-LSCO, or 
two-dimensional ``patches'' in Bi2212. In the former case, the inter-stripe spacing in the 
superconducting regime is reported to be approximately  four times the in-plane lattice 
constant, about 1.5 nm, and in the latter case the patches  are estimated to be 1-3 nm across.
 Many theoretical studies suggest that the spatial electronic inhomogeneity in the hole-doped 
CuO$_2$ planes is an essential part of high-\Tc~physics~\cite{Theory}. 
However, at present, the importance of, or even the existence of generic
 nanoscale electronic inhomogeneity remains controversial~\cite{controversy,bobroff}.

If nanoscale electronic inhomogeneity exists in the superconducting cuprates, the doped holes 
will distribute themselves in the CuO$_2$ planes so as to minimize their total energy. 
In real materials, the CuO$_2$ planes are usually inhomogeneous due to local lattice distortions
 and/or the random Coulomb potential resulting from chemical disorder, which differs from system
 to system.
Therefore, even if electronic inhomogeneity may itself be a genuine property of doped 
CuO$_2$ planes, the spatial variation of doped holes will likely depend on the details of 
each material. For example, in the framework of the stripe model~\cite{neutron}, incommensurate 
spin and charge correlations are stabilized in Nd-LSCO by the long-range distortion of the 
CuO$_6$ octahedra in the low-temperature tetragonal phase, which creates one-dimensional potential wells. 
For Bi2212, it is argued that the random Coulomb potential caused by excess oxygen atoms in the 
BiO planes pins the doped holes, thus creating patch-shaped inhomogeneities~\cite{STM}. 
These observations suggest that {\it electronic and chemical inhomogeneity are inseparable 
from each other}, and that the understanding of the latter is imperative for an understanding 
of the former.

Motivated by this line of reasoning, we have examined the effects of chemical inhomogeneity in
 single-layer Bi$_2$Sr$_2$CuO$_{6+ \delta}$ (Bi2201) and double-layer Bi2212. 
Although widely used for surface sensitive measurements such as STM~\cite{STM}
 and angle-resolved photoemission spectroscopy (ARPES)~\cite{Andrea},
a detailed understanding of their materials properties is 
very limited, when compared to other materials such as LSCO or YBa$_2$Cu$_3$O$_{7-\delta}$.

The Bi-based cuprates contain excess oxygens in BiO planes and one can change 
their carrier concentration by changing the amount, $\delta$. The excess oxygen
 would engender a random Coulomb potential in the CuO$_2$ planes.  
Besides the oxygen nonstoichiometry in BiO planes, there exists another source of chemical 
inhomogeneity which inherently exists in typical samples.
Although referred to as Bi2201 and Bi2212, it is empirically known that 
stoichiometric Bi$_{2.0}$Sr$_{2.0}$CuO$_{6+\delta}$ 
and Bi$_{2.0}$Sr$_{2.0}$CaCu$_2$O$_{8+\delta}$ are very difficult to
synthesize~\cite{Maeda, lone_pair}, even in a polycrystalline form. 
In order to more easily form the crystal
structure, one usually replaces Sr$^{2+}$ ions by trivalent ions, such as excess
 Bi$^{3+}$ ions or La$^{3+}$ ions, forming Bi$_{2+x}$Sr$_{2-x}$CuO$_{6+\delta}$,
Bi$_2$Sr$_{2-x}$La$_{x}$CuO$_{6+\delta}$, and Bi$_{2+x}$Sr$_{2-x}$CaCu$_2$O$_{8+\delta}$. 
As listed up in Ref.~\onlinecite{Bi}, a typical Bi:Sr nonstoichiometry, $x$, for Bi2212 is
 around 0.1, which yields a $T_c$=89-91K. To our knowledge, the highest
$T_c$ reported in the literature is 95K (Ref.\onlinecite{Bi} (g),(i)). For Bi2201, $T_c$ of
Bi$_{2+x}$Sr$_{2-x}$CuO$_{6+\delta}$ is around 10K for $x$(Bi)$=0.1$, 
whereas La substituted Bi2201 (Bi$_2$Sr$_{2-x}$La$_{x}$CuO$_{6+\delta}$) has a higher 
$T_c>$30K for $x$(La)$\approx0.4$~\cite{ando}.
Since the Sr atom is located next to the apical oxygen which is just above the Cu atoms,
 the effect of Sr site (also referred to as the A-site) cation inhomogeneity is
 expected to be stronger than that of the excess oxygens in BiO planes. Note that BiO planes
are located relatively far away from CuO$_2$ planes, with SrO planes in between.

In this study, we evaluate the effect of chemical inhomogeniety in the Bi-based
cuprates. For Bi2201, we have grown a series of Bi$_2$Sr$_{1.6}$Ln$_{0.4}$CuO$_{6+\delta}$ 
crystals with various trivalent rare earth (Ln) ions. 
In this series, the magnitude of the local lattice distortion can be changed systematically
 by making use of the different ionic radii of the substituted Ln ions. We find that 
\Tc~ monotonically decreases with increasing ionic radius mismatch. For Bi2212,
a series of Bi$_{2+x}$Sr$_{2-x}$CaCu$_2$O$_{8+\delta}$ crystals
 with varying values of $x$ were grown in order to evaluate the effect of Bi:Sr nonstoichiometry. 
In addition, we also have grown Bi$_2$Sr$_2$Ca$_{1-y}$Y$_y$Cu$_2$O$_{8+\delta}$, and 
 find that  substitution of Y for Ca site helps to enforce Bi:Sr stoichiometry and to 
raise \Tc~to 96 K for $y=0.08$.

Our results demonstrate that the cation disorder, in particular that located at the Sr 
site, significantly affects the maximum attainable \Tc (\Tcm)~in the Bi-based superconductors. 
In order to explain our results we use a conceptual hierarchy that classifies and ranks 
the principal kinds of chemical disorder possible in these systems. We then extend our 
arguments to other cuprates to examine whether a general trend exists in the hole-doped
 high-$T_c$ superconductors.

This paper is organized as follows: Section II contains detailed information about sample 
preparation and characterization. The experimental results are presented in Section
 III and discussed in Section IV, while the effects of disorder in other cuprates is
 addressed in Section V.

\section{SAMPLE PREPARATION}

Single crystals of Bi2201 and Bi2212 were grown using the travelling-solvent floating-zone 
technique, which is now the preferred method for synthesizing high-purity single crystals 
of many transition metal oxides. This technique allows for greater control of the growth 
conditions than is possible either by standard solid state reactions or by the flux method. 

Powders of Bi$_2$O$_3$, SrCO$_3$, CaCO$_3$, Ln$_2$O$_3$ (Ln=La, Pr, Nd, Sm, Eu, Gd, Y), and 
CuO (all of 99.99\% or higher purity) were well dried and mixed in the desired cation ratio 
 (Bi:Sr:Ln:Cu=$2:2-x:x:1$ for Bi2201 and Bi:Sr:Ca:Cu=$2+x:2-x:1:2$ for Bi2212), 
and then repeatedly calcinated at about 800$^\circ$C with intermediate grinding. 
Eventually, the powder was finely ground and formed into a 100 mm long rod with 
a diameter of 5 mm. The crystal growth was performed using a Crystal Systems Inc. 
infrared radiation furnace equipped with four 150 W halogen lamps. 
Except for nearly-stoichiometric ($x=0$) Bi2212, the rods were premelted at 18 mm/h 
to form dense feed rods. The crystal growth was carried out without the use of a 
solvent and at a growth speed of 0.3-0.4 mm/h for Bi2201 and 0.15-0.2 mm/h for Bi2212. 
The growth atmospheres adopted for the Bi2212 growth are listed in 
Table 1. Bi2201 single crystals were grown in 1 atm of flowing O$_2$. 

The growth condition for the $x=0$ Bi2212 sample was more stringent than for
 the other samples. In order to obtain homogeneous polycrystalline feed rods, 
a mixture of starting powders with the stoichiometric ratio Bi:Sr:Ca:Cu=2:2:1:2  
was calcinated in stages, at temperatures increasing from 770$^\circ$C to 870$^\circ$C in
 10$^\circ$C increments, with intermediate grindings. The final calcination temperature 
(870$^\circ$C) was set to be just below the composition's melting temperature (875$^\circ$C). 
The duration of each calcination was about 20 hours. 
To avoid possible compositional fluctuation in the feed rod, instead of premelting, 
the feed rod was sintered four times in the floating-zone furnace at a speed of 50 mm/h. 
This process allowed us to obtain dense feed rods, 
approximately 95\% of the ideal density. 
The atmosphere required for stable crystal growth was $7\pm\;3\%$ O$_2$, 
a range much narrower than for nonstoichiometric or Y-doped Bi2212. 
The grown crystal rod contained small amounts of a single-crystalline SrCuO$_2$ 
secondary phase, indicating that the sample still suffers from 
Bi:Sr nonstoichiometry. Single-phase Bi2212 single crystals could be cleaved from 
the grown rod. No traces of impurity phases were found in other compositions.

Inductively coupled plasma (ICP) spectroscopy was carried out to determine the chemical 
compositions of the crystals. Additional electron-probe microanalysis (EPMA) was also carried 
out on the Bi2201 crystals. The results confirm that the actual compositions follow the nominal
 compositions, as listed in Table 1 for Bi2212.  Hereafter, we basically denote the samples by 
nominal composition to avoid confusion. In order to determine the maximum \Tc~for each cation
 composition (i.e., to achieve optimal hole concentration), the Bi2212 samples were annealed at 
various temperatures and oxygen partial pressures using a tube furnace equipped with an 
oxygen monitor and a sample transfer arm, which allowed us to rapidly quench the annealed 
samples from high temperatures within a closed environment. This procedure ensures that the
 variation of \Tc~among the samples is primarily due to cation nonstoichiometry and not due
 to differing hole concentrations. Annealing conditions for obtaining optimal (OP), and 
typical underdoped (UD) and overdoped (OD) samples are listed in Table 1. The results 
 for Bi2201 are on the as-grown crystals and accordingly may not exactly reflect \Tcm. 
However, by carrying out a series of annealing studies, we have confirmed that the 
systematic change of \Tc~among our samples is not due to different hole concentrations, 
but due to the different Ln ions.
 
\begin{table*}[t]
\caption{\label{tab:table1}Sample preparation conditions and crystal compositions derived 
from the ICP analysis for Bi2212 single crystals. The ICP analysis on the third sample in the Table
was done on cleaved, single-phase samples from an ingot containing a 
small amount of SrCuO$_2$ secondary phase.}
\begin{ruledtabular}
\begin{tabular}{ccccccc}
 &&&&\multicolumn{3}{c}{annealing condition}\\
 nominal composition&ICP results&Bi:Sr ratio&growth atmosphere&UD&OP&OD\\ \hline
 Bi$_{2.2}$Sr$_{1.8}$CaCu$_2$O$_{8+\delta}$&Bi$_{2.19}$Sr$_{1.86}$Ca$_{1.07}$Cu$_2$O$_{8+\delta}$&1.178 & 1atm & 700$^\circ$C & 750$^\circ$C & 400$^\circ$C  \\ 
 & & & O$_2$ & air & O$_2$ & O$_2$ \\
 Bi$_{2.04}$Sr$_{1.96}$CaCu$_2$O$_{8+\delta}$&Bi$_{2.06}$Sr$_{1.93}$Ca$_{0.96}$Cu$_2$O$_{8+\delta}$&1.069 & 3atm &650$^\circ$C & 500$^\circ$C & as-grown \\ 
 & & & O$_2$ & Ar & Ar &  \\
 Bi$_{2.00}$Sr$_{2.00}$CaCu$_2$O$_{8+\delta}$&Bi$_{2.06}$Sr$_{2.04}$Ca$_{0.87}$Cu$_2$O$_{8+\delta}$ &1.008 & 1atm & 650$^\circ$C & 500$^\circ$C & 650$^\circ$C \\
 & & & O$_2$:Ar=7:93 & Ar & Ar & O$_2$ \\
 Bi$_{2.00}$Sr$_{2.00}$Ca$_{0.92}$Y$_{0.08}$Cu$_2$O$_{8+\delta}$&Bi$_{2.02}$Sr$_{2.01}$Ca$_{0.85}$Y$_{0.08}$Cu$_2$O$_{8+\delta}$ & 1.004 & 1atm & 650$^\circ$C & 400$^\circ$C  & 500$^\circ$C \\
 & & & O$_2$:Ar=20:80 & Ar & Ar & O$_2$ \\

\end{tabular}
\end{ruledtabular}
\end{table*}

Superconducting transition temperatures were determined by AC susceptibility measurements using 
a Quantum Design Physical Properties Measurement System (PPMS). The transition temperatures 
reported here correspond to the onset of a diamagnetic signal. We note that the different 
definition of \Tc (such as the interecept between the superconducting transition slope and 
the $\chi=0$ axis) does not affect our conclusions due to the sharp 
superconducting transition (less than 2K for most samples), as shown below.
Although it is hard to determine the exact superconducting fraction due to the demagnetization factor of the plate-shaped crystals, 
the magnitude of the superconducting signal suggests the bulk superconductivity of the grown samples.
No appreciable differences were observed between different samples from the same growth, 
or between different crystals prepared under identical conditions, an indication of the 
macroscopic homogeneity of the crystals and the reproducibility 
of our sample growth process. 

\section{EXPERIMENTAL RESULTS}

\subsection{Results for Bi2201}

We have grown a series of Ln$^{3+}$ substituted Bi$_2$Sr$_{1.6}$Ln$_{0.4}$CuO$_{6+\delta}$ 
single crystals with Ln = La, Pr, Nd, Sm, Eu, Gd, Bi. The ionic radius of Ln$^{3+}$ ions, 
$R_{Ln}$, decreases monotonically with increasing atomic number, 1.14 \AA ~(La$^{3+}$), 
1.06 \AA ~(Pr$^{3+}$), 1.04 \AA ~(Nd$^{3+}$), 1.00 \AA ~(Sm$^{3+}$), 0.98 \AA ~(Eu$^{3+}$), 
0.97 \AA ~(Gd$^{3+}$), and 0.96 \AA ~(Bi$^{3+}$)\cite{ionic_radius}. 
Accordingly, by introducing different Ln ions, we can systematically change the ionic radius 
mismatch between Sr$^{2+}$ (1.12 \AA) and Ln$^{3+}$. We use this mismatch, defined as 
$\Delta R \equiv|R_{Sr} - R_{Ln}|$, to quantify the magnitude of the local lattice distortion 
around Ln$^{3+}$ ions.

\begin{figure}
 \includegraphics[width=8cm,clip]{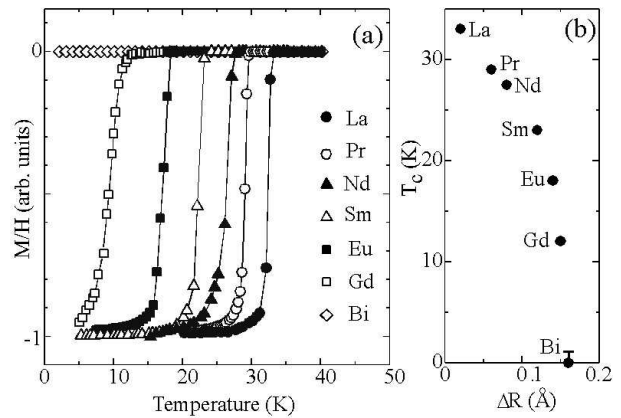}
\caption{\label{fig:epsart}(a) Bi$_2$Sr$_{1.6}$Ln$_{0.4}$CuO$_{6+\delta}$ 
susceptibility curves,
normalized to -1 at the 
lowest temperature.
(b) $T_c$ values as a function of $\Delta R$. }
 \end{figure}

In Fig. 1(a), we show magnetic susceptibility data for a series of Ln$^{3+}$ substituted
 samples grown under the same conditions. The \Tc~of the La-doped sample was 33 K
 with a transition width of less than 1 K, a reasonable result for an as-grown crystal
 with this composition~\cite{ando}. As shown in Fig. 1(b), $T_c$ decreases monotonically 
with increasing $\Delta R$, yielding 29 K (Pr$^{3+}$), 27.5 K (Nd$^{3+}$), 23 K (Sm$^{3+}$), 
18 K (Eu$^{3+}$), and 12 K (Gd$^{3+}$), respectively.  
There is no trace of superconductivity down to 1.8 K for the Bi-substituted sample. 
This is reasonable, since the ionic radius of 
Bi$^{3+}$ is the smallest of the Ln ions, and the shape of the Bi$^{3+}$ ion tends to be asymmetric 
due to the presence of a (6s)$^2$ lone pair~\cite{lone_pair}, 
which might cause additional local lattice distortions. 
These observations demonstrate the strong sensitivity of \Tc~to Sr site substitution in Bi2201.

Although one can minimize the magnitude of the local lattice distortion through the 
choice of the La$^{3+}$ ion, which has a radius similar to that of Sr$^{2+}$,
there remains another type of chemical disorder, that is, the random Coulomb
potential caused by heterovalent ion substitution. 
For LSCO, $^{63}$Cu NQR spin-lattice relaxation rates experiments have revealed 
signatures for an electronic inhomogeniety and the results have been 
discussed in connection with the random distribution of Sr$^{2+}$ ions\cite{dirt}. 
Considering that both LSCO and Bi2201 contain A-site chemical disorder,
 it seems likely that the disorder would affect the electronic properties 
of Bi2201 to a degree comparable to that observed in LSCO, and much more strongly 
than for the structurally similar material
Tl$_2$Ba$_2$CuO$_{6+\delta}$ (Tl2201) which is believed to be essentially free of 
A-site disorder.
We will discuss this issue in more detail below.

\subsection{Results for Bi2212}

Since cation substitution at the Sr site has such a dramatic effect on \Tc~in Bi2201, 
in particular for Ln=Bi, one might expect to find similar results for Bi2212. 
To verify this, we have grown single crystals of Bi$_{2+x}$Sr$_{2-x}$CaCu$_2$O$_{8+\delta}$. 
By adopting the methods described in Sec. II,
 we have managed to grow single crystals over the range $0.0<x\le0.2$. 
In Figs. 2(a)-(c), we present magnetic susceptibility data for three different crystals with 
 compositions $x = 0.2$, 0.04, and $\simeq$ 0, respectively ($\epsilon$ in the chemical formula 
for the nominal $x = 0$ sample (Fig. 2 (c)) implies the presence of residual nonstoichiometry in 
our sample, as discussed in the previous section). 
 In the figures, OP indicates optimally-doped samples, 
which possess \Tcm~for a given cation composition. 
Representative data for underdoped (UD) and overdoped (OD) samples, 
which were obtained by reducing and oxidizing OP samples, are also plotted 
to demonstrate successful control of the hole concentration over a wide range.

As the Bi:Sr ratio approaches 1:1, \Tcm~increases from 82.4 K for $x=0.2$, to 91.4 K 
for $x=0.07$ (not shown), 92.6 K for $x=0.04$, and eventually to 94.0 K for the sample 
closest to the stoichiometric composition that we could grow.
We note that most of the 
samples studied in the literature contain a nonstoichiometry of $x\sim0.1$ with 
$T_{\rm c} = 89-91$~K\cite{Bi}, consistent with the present results. 

\begin{figure}

\includegraphics[width=7cm]{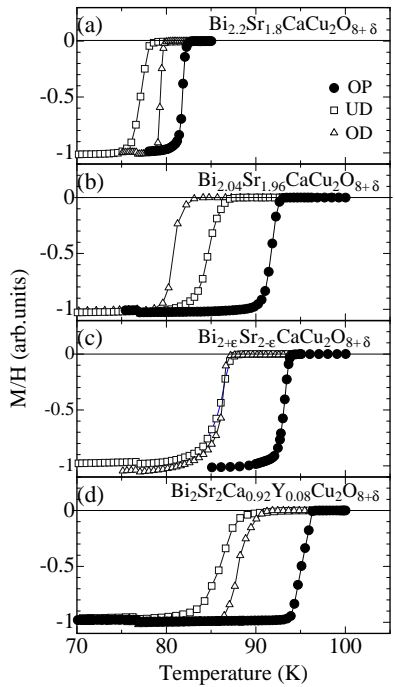}
\caption{ Bi$_{2+x}$Sr$_{2-x}$Ca$_{1-y}$Y$_y$Cu$_2$O$_{8+\delta}$
susceptibility curves,
normalized to -1 at the lowest temperature.
 Data for optimally-doped (OP), underdoped (UD), and overdoped 
(OD) samples are indicated for each cation composition 
by closed circles, open squares, and open triangles, respectively. }
\end{figure}

Although  \Tc~of Bi2212 can be raised by trying to enforce
Bi:Sr stoichiometry, the preparation of nearly stoichiometric samples
becomes much more difficult than when nonstoichiometry is allowed. This
could be due to a greater stability of the crystal structure when
it contains additional positive charges, which are usually introduced 
by allowing the Bi$^{3+}$:Sr$^{2+}$ ratio to be larger than one, as discussed in 
Ref. \onlinecite{lone_pair}. 
If this is indeed the case, one might expect to be able to synthesize 
higher-\Tcm~samples more easily by introducing
extra positive charges via cation substitution that
causes disorder less severe than substitution of Bi$^{3+}$ ions at the Sr site.

In the case of Bi2201 we observed that the substitution of additional Ln atoms 
can eliminate excess Bi atoms from the unfavorable Sr site position, effectively
 lowering the magnitude of disorder and raising \Tc. In the double-layer material
 Bi2212, there is an additional crystallographic site, the Ca site located between
 the CuO$_2$ planes, which can also accept  trivalent dopant ions. One might expect
 Ln$^{3+}$ ions at the Ca site to be a weaker type of disorder than Bi$^{3+}$ ions at the 
Sr site, since there are no apical oxygens in the Ca planes that could couple to Cu atoms 
in the CuO$_2$ planes.

To test this idea we have also grown Y-substituted Bi$_2$Sr$_2$Ca$_{1-y}$Y$_y$Cu$_2$O$_{8+\delta}$ crystals. 
We find that this compound is as easy to prepare as ordinary (nonstoichiometric) Bi2212, 
and that for Y-Bi2212 the Bi:Sr ratio indeed tends to be stoichiometric. Furthermore, as shown
 in Fig. 2(d), \Tcm~for the $y=0.08$ sample was increased to 96.0 K, a value higher than for any
other Bi2212 sample reported in the literature.\cite{Bi} We also grew $y=0.10$ and $y=0.12$ 
samples and confirmed that \Tcm $> 95$ K in both cases.

\section{DISCUSSION}

The effect on \Tc~of structural distortions associated with cation substitution has been 
extensively studied in LSCO-based materials~\cite{buckling}, and it is established that \Tc~ 
strongly depends on the A-site (La site) ionic radius mismatch.
For instance, Attfield {\it et al.} used simultaneous co-substitution of
several alkaline earth and Ln ions to hold the average A-site ionic radius 
constant while systematically controlling the variance of the A-site ionic radius, and found that 
\Tc~is affected not just by the average radius, but also by the degree of disorder (the 
variance) at that crystallographic site. Our study of single-layer Bi2201 continues
 this line of inquiry to a different superconducting material and demonstrates a similar
 sensitivity of \Tc~to A-site disorder. 
We note that our results qualitatively agree with  
those of a previous study on polycrystalline Bi2201 samples\cite{syono}.

One can see the same trend in Bi2212 crystals with varying degrees of chemical inhomogeniety. 
As expected, we find that \Tc~is strongly dependent on the A-site disorder introduced by
the Bi:Sr nonstoichiometry. 
Furthermore, we also demonstrate that by the seemingly counter-intuitive method of 
introducing additional Y$^{3+}$ ions, and hence a new type of disorder, 
we can raise \Tcm~to 96 K while minimizing A-site disorder. 
This suggests that, although the minimization of chemical disorder is important for 
raising \Tc, different types of disorder are not equally harmful. 
This is consistent with the observation
\cite{Iyo} that by carefully controlling disorder in the triple-layer material 
TlBa$_2$Ca$_2$Cu$_3$O$_{9+\delta}$ (Tl1223) $T_c$ can be raised from $\sim120$ K to 133.5 K, 
a new record for that system, and that Ba site (the A-site in this system) cation disorder(deficiency)
 has the strongest effect on \Tc.

Numerous experiments on Bi2212 have suggested non-uniformity in its electronic properties. 
These include broad linewidths seen in inelastic neutron scattering experiments~\cite{Keimer}, 
residual low-energy excitations in the superconducting state observed in penetration 
depth measurements~\cite{penetration}, finite spin susceptibility at low temperatures observed 
in NMR studies~\cite{Takigawa}, and short quasiparticle lifetimes detected by complex conductivity
 experiments~\cite{Orenstein}. 
The most recent of these are STM/STS measurements \cite{STM} that purport to directly
 image patch-shaped, electronically inhomogeneous regions. 
The Bi:Sr nonstoichiometry which inherently exists in most samples may be partially responsible
 for these experimental observations.

Although the present results do not directly prove the presence of nanoscale 
electronic inhomogeneity, they can be taken as a circumstantial supporting 
evidence, since they successfully prove the existence of nanoscale chemical inhomogeneity 
which potentially pins down electronic inhomogeneity.  
To explain their STM/STS results, Pan {\it et al.}\cite{STM} attribute the source of 
pinning centers to excess oxygen in the BiO planes. 
Although the overall framework addressed by Pan {\it et al.} should still hold, 
we consider that the Bi ions on the Sr site are more effective as pinning centers
 since they are closer to the CuO$_2$ planes and affect \Tc~more directly. 
Indeed, assuming a random distribution of Bi ions on the Sr site and a nonstoichiometry
 of $x=0.10$, the average separation between Bi ions is
 $\sim 1-2$ nm, comparable with the length scale observed in the STM/STS studies.  

Recent $^{89}$Y NMR experiments on YBCO indicate that the spatial inhomogeniety in this 
system is much less severe than in LSCO or Bi2212~\cite{bobroff}. 
This is reasonable since the latter two systems exhibit a much higher degree of 
disorder located at the A-site (La site (LSCO) and Sr site (Bi2212)), 
whereas YBCO (Ba site) is thought to be free from such cation disorder. 
Indeed, recent penetration depth measurements on
the YBCO variant Nd$_{1+x}$Ba$_{2-x}$Cu$_3$O$_{7-\delta}$, with cationic disorder
at the Ba site,  demonstrate
that the superconducting properties of this system
change quite sensitively with the degree of Nd/Ba nonstoichiometry~\cite{salluzzo}.

\section{DISORDER EFFECTS IN THE CUPRATES}

\begin{figure*}[t]

\includegraphics[width=15cm]{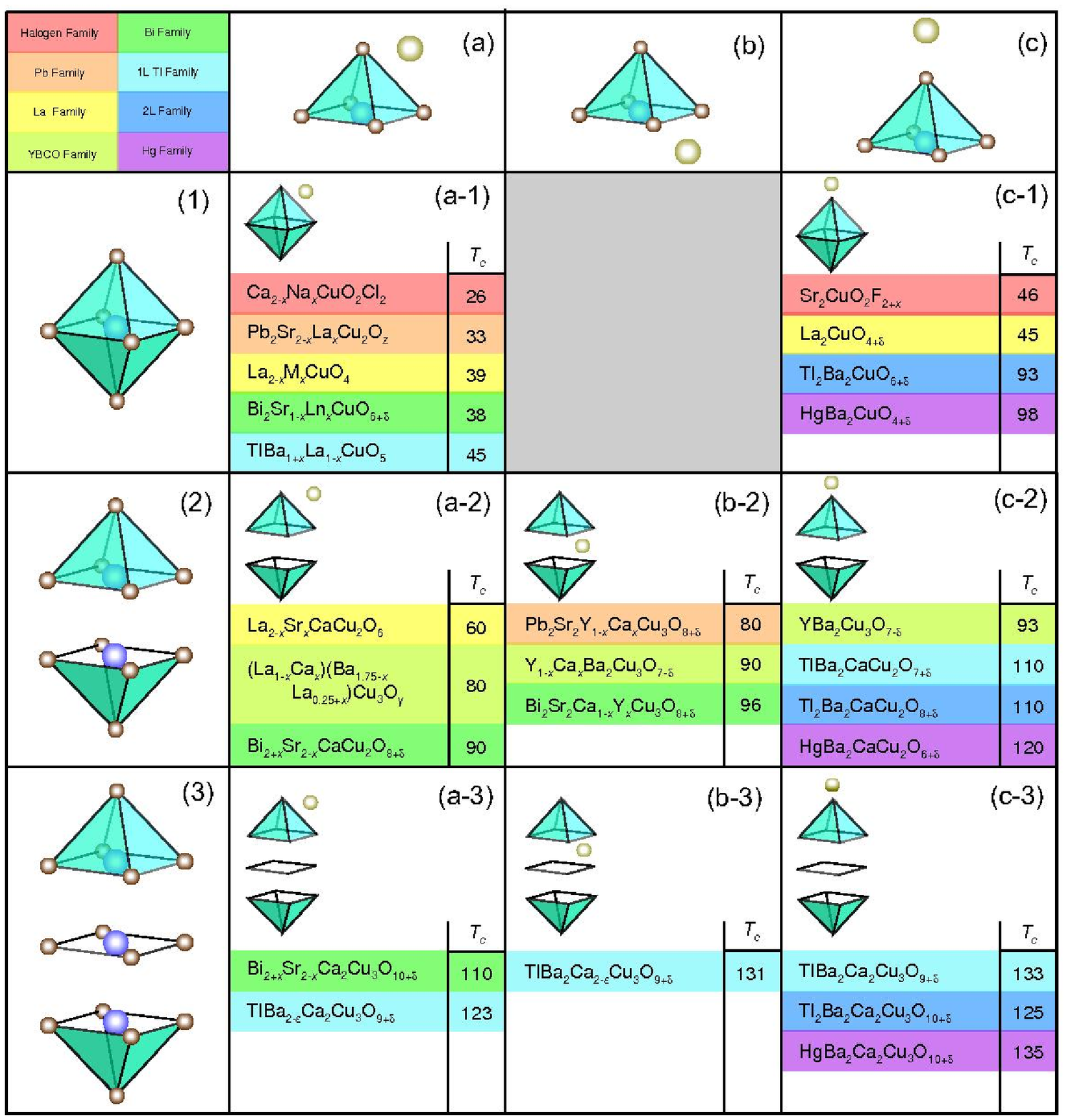}
\caption{ (Color) Classification of bulk high-$T_c$ cuprates in
terms of the disorder site and the number of CuO$_2$ layers.
Materials belonging to the same family are indicated by the same
color. Halogen family denotes (Ca,Sr)$_2$CuO$_2$A$_2$ (A=Cl, F)
based materials. Bi family denotes Bi22(n-1)n (n=1,2,3). Pb family
denotes Pb$_2$Sr$_2$Ca$_{n-1}$Cu$_{n+1}$O$_z$.  1L Tl family
denotes one-Tl-layer cuprates,
TlBa$_2$Ca$_{n-1}$Cu$_n$O$_{3+2n+\delta}$. 2L Tl family denotes
two-Tl-layer cuprates,
Tl$_2$Ba$_2$Ca$_{n-1}$Cu$_n$O$_{4+2n+\delta}$. La family denotes
La$_2$Ca$_{n-1}$Cu$_n$O$_{2+2n}$.  YBCO family denotes
LnBa$_2$Cu$_3$O$_{6+\delta}$. Hg family denotes
HgBa$_2$Ca$_{n-1}$Cu$_n$O$_{2+2n+\delta}$. The transition temperatures 
are compiled from works listed in Ref. \onlinecite{zoology} }

\end{figure*}

The two main lessons to be learned from our Bi2201 and Bi2212 case studies are that
 (1) chemical inhomogeneity affects \Tcm~and that 
(2) the effect of disorder differs depending on its location. 
In the following, we attempt to classify the various sites 
at which chemical disorder is possible and categorize other superconducting families 
on the basis of which kind of disorder is prevalent in each system.

In Fig. 3, we classify 25 cuprate superconductors based on the pattern
 of the chemical disorder and the number of CuO$_2$ planes in the unit cell. \cite{zoology} 
In the first row, we illustrate three possible locations of chemical disorder relative to the 
CuO$_5$ pyramids in multilayer materials, or to the CuO$_6$ octahedra in single-layer materials. 
Pattern (a) corresponds to the Bi:Sr nonstoichiometry in Bi2201 and Bi2212, or Sr$^{2+}$ ions
 doped into the La site in LSCO, referred to as A-site disorder so far. The disorder is located
 next to the apical oxygen. Pattern (b) corresponds to Y$^{3+}$ substitution for Ca$^{2+}$ in 
Bi2212 and represents disorder located next to the CuO$_2$ plane, but at a position where there 
are no apical oxygen atoms with which to bond. There is no corresponding (b) site in single-layer 
materials. Pattern (c) disorder is further away from the CuO$_2$ plane. We include excess oxygen 
$\delta$ in Bi- and Tl-based cuprates, oxygen defects in CuO chains in YBa$_2$Cu$_3$O$_{7-\delta}$, 
and Hg deficiency $y$ as well as excess oxygen $\delta$ in 
Hg$_{1-y}$Ba$_2$Ca$_{n-1}$Cu$_n$O$_{2n+2+\delta}$ in this category. 
We note that the materials are catalogued based on the $primary$ form of disorder that they 
are believed to exhibit.

As demonstrated in the present case study, the effect of the chemical disorder is 
expected to be stronger for pattern (a) than for pattern (b), which is reasonable 
considering the role of the apical oxygen atom in passing on the effect of disorder 
to nearby Cu atoms. First, the random Coulomb potential caused by type (a) disorder 
changes the energy levels of the apical oxygen orbitals, which can be transmitted 
to CuO$_2$ planes through the hybridization between the apical O(2p$_z$) orbital and 
the Cu(3d$_{r^2- 3z^2}$) orbital. Second, the displacement of the apical oxygen caused 
by type (a) disorder brings about a local lattice distortion to the CuO$_2$ planes. 
The effect of pattern (c) disorder is expected 
to be weakest since the disorder is located relatively far away from the CuO$_2$ plane.

The number of CuO$_2$ planes per unit cell may be regarded as another parameter that, 
in effect, determines the magnitude of the chemical disorder. As demonstrated in Bi2212, 
multilayer materials can accommodate heterovalent ions by making use of type (b) substitution 
whose effect on \Tc~was seen to be weaker than type (a) substitution. Furthermore, the space 
between the CuO$_2$ planes forming a multilayer may buffer the impact of  the disorder. 
For instance, in single-layer materials, any displacement of the ``upper" apical oxygen in 
a CuO$_6$ octahedron creates stress in the CuO$_2$ plane because the motion of the octahedron 
is constrained by the ``lower" apical oxygen. Double-layer materials contain CuO$_5$ pyramids 
rather than CuO$_6$ octahedra, and the separation between CuO$_2$ planes relieves this stress, 
reducing the effects of type (a) disorder. This buffer zone between the outer CuO$_2$ planes is 
further increased in triple-layer materials, with the additional benefit that the middle layer 
is somewhat ``protected" from the direct effects of pattern (a) (and (c)) disorder.

Cursory examination of Fig. 3 reveals that \Tcm~generally increases both across the rows and 
down the columns of the chart. Indeed, there is no material in (a-1) which possesses a 
\Tcm~higher than 50 K. Furthermore, Bi2201 in column (a) has a lower \Tcm~than Tl2201 in 
column (c), despite their similar crystal structures. Similarly, \Tcm of 
TlBa$_{1+x}$La$_{1-x}$CuO$_5$ (Tl1201) is lower than that of HgBa$_2$CuO$_{4+\delta}$ (Hg1201). 

This trend is closely obeyed when 
one concentrates on the variation within a single family of materials, 
each denoted by a different color in the chart. For example, across the first row, 
oxygen-intercalated La$_2$CuO$_{4+\delta}$ located in (c-1) has higher \Tcm~than 
Sr-substituted La$_2$CuO$_4$ (a-1). Down the column, \Tcm~of the bilayer La-based 
system La$_{2-x}$Sr$_x$CaCu$_2$O$_6$ (a-2) is higher than that of its single-layer cousin. 
The classification suggests a negative
correlation between the effective magnitude of chemical disorder and \Tcm.
Additional remarks are made in Ref.s \onlinecite{remark2, remark3, remark4}. 
 
We note that one may also have to consider other 
factors which characterize the {\it global} materials 
properties and are likely to play a significant role as well in determining \Tcm,
 such as Madelung potential\cite{Madelung}, 
bond valence sum\cite{bvs}, band structure \cite{band}, block layer \cite{blocklayer}, 
multi layer \cite{multilayer} etc. 
Although the present scheme is somewhat oversimplified and does not take account of
these parameters, we believe it serves as a useful framework within which to 
consider the chemical disorder effects prevalent in these materials, 
at least some of which, if ignored, have the potential to lead to the 
misinterpretation of experimental data. 
Finally, similar to previous work on Tl1223 \cite{Iyo} and to the present work on 
Bi2212, it might be possible to raise \Tcm~of certain other 
cuprates by minimizing the effects of chemical disorder.

\indent

\section{SUMMARY}

In summary, we present case studies of the effects of chemical disorder on 
the superconducting transition temperature of the single-layer and double-layer 
Bi-based cuprate superconductors. 
We find that the superconducting transition temperature of Bi2212 can be 
increased up to 96 K by lowering the impact of Sr site disorder, 
the primary type of disorder inherent to the bismuth family
 of materials, at the expense of Ca site disorder. 
Based on these experimental results, we present a qualitative 
hierarchy of possible disorder sites, and then proceed to categorize the
hole-doped high-temperature superconductors on that basis. 

\begin{acknowledgments}

We thank G. Blumberg, J. Burgy, E. Dagotto, J. C. Davis, D. S. Dessau, 
T. H. Geballe, E. W. Hudson, A. Iyo, A. Kapitulnik, 
G. Kinoda, S. A. Kivelson, R. B. Laughlin, B. Y. Moyzhes, D. J. Scalapino, 
T. Timusk, and J. M. Tranquada for helpful discussions. 
This work was supported by the NEDO grant ``Nanoscale phenomena of 
self-organized electrons in complex oxides-new frontiers in physics 
and devices", by the U.S. DOE under contract 
nos. DE-FG03-99ER45773 and DE-AC-03-76SF00515, by NSF DMR9985067, DMR0071897, 
ONR N00014-98-0195 
and by the DOE's Office of Basic Energy Science, Division of Material Science, 
Divison of Chemical Science, through the Stanford Synchrotron Radiation Laboratory (SSRL). 
H. Eisaki was supported by the Marvin Chodorow Fellowship in 
the Department of Applied Physics, Stanford University.

\end{acknowledgments}

\end{document}